\documentclass[prl,aps,twocolumn]{revtex4-1}
\usepackage[latin1]{inputenc}
\usepackage[english]{babel}
\usepackage{graphicx}
\usepackage{color}
\usepackage{amsmath}
\usepackage{amssymb}

\begin{document}

\title{\bf  The Generalized Schr\"odinger-Langevin equation}
\author{Pedro Bargue\~no$^{1}$ and Salvador Miret--Art\'es$^{2}$}
\affiliation{$^{1}$ Departamento de F\'{\i}sica, Universidad de los Andes, Apartado A\'ereo 4976, Bogot\'a, Distrito Capital, Colombia
(p.bargueno@uniandes.edu.co)
\\
$^{2}$ Instituto de F\'{\i}sica Fundamental, CSIC, Serrano 123,
{\it 28006}, Madrid, Spain (s.miret@iff.csic.es)
}

\begin{abstract}
In this work, we derive a generalization of the so-called
Schr\"odinger-Langevin or Kostin equation for a Brownian particle
interacting with a heat bath. This generalization is based on a
nonlinear interaction model providing a state-dependent
dissipation process exhibiting multiplicative noise. Two
straightforward applications to the measurement process are then
analyzed, continuous and weak measurements in terms of the quantum
Bohmian trajectory formalism. Finally, it is also shown that the
generalized uncertainty principle, which appears in some
approaches to quantum gravity, can be expressed in terms of this
generalized equation.

\end{abstract}

\maketitle


Quantum stochasticity constitutes a very broad and active field of
research within quantum mechanics. Real physical systems do not
exist in complete isolation and one then speaks about open quantum
systems \cite{Weiss1999,Petruccione2002,salva2013}. Thus, the
interaction of a quantum system  with its environment can not be
totally neglected leading to an entanglement between them. The
corresponding theory encompasses a series of formalisms and
approaches developed to deal with this complex but fundamental
issue. Three main approaches are usually considered in this
context: (i) effective time-dependent Hamiltonians, (ii) nonlinear
(logarithmic) Schr\"odinger equations and (iii) the
system-plus-bath model within a conservative scenario. Obviously,
links among them can be found. For example, in the last approach,
and for one dimensional systems, the so--called Caldeira-Leggett
Hamiltonian \cite{Caldeira1981} is the starting point leading to
the generalized Langevin equation (GLE). One of the key issues is
the interaction term which by construction is linear in the bath
coordinates. The dependence on the system variable is through a
function $f(x)$ which usually is separable and linear (linear
dissipation). This scenario is known as a state--independent
dissipation and can be seen as a measurement of particle's
position by a reservoir in von Neumann's sense. \cite{Weiss1999}
This function also appears in an additional term in the total
Hamiltonian in order to avoid the renormalization of the
interaction potential. In the Markovian regime with Ohmic
friction, this GLE becomes the standard Langevin equation for a
Brownian particle where noise is additive. Kostin
\cite{Kostin1972} established the link between this standard
Langevin equation with the Schr\"odinger equation leading to a
nonlinear, logarithmic equation termed the Schr\"odinger-Langevin
(SL) or Kostin equation. This type of equations are quite
different from others existing in the literature such as the
so--called stochastic Schr\"odinger equation and the Linblad
equation for the density matrix. \cite{Petruccione2002}

For a nonlinear function $f(x)$, the open quantum system displays
a state dependent dissipation process and the corresponding GLE
exhibits multiplicative noise. \cite{Hanggi1995,Sancho1982}
Typical examples of nonlinear functions take place, for example,
in rotational tunneling systems \cite{Stevens1983}, quasi-particle
tunneling in Josephson systems \cite{Ambegaokar1987}, in the
Langevin canonical formulation of chiral two level systems
\cite{Anais2012}, in atom surface scattering \cite{Pollak2012} and
so on. Within the Markovian regime, the standard Langevin equation
with multiplicative noise is reached. The main purpose of this
work is to derive a generalization of the SL or Kostin equation
for nonlinear dissipation which is termed the generalized SL
equation (GSLE).

Once this equation is established, three straightforward
applications dealing with two measurement processes and quantum
gravity are analyzed. The first one considers continuous
measurement. As is known, the mere presence of an observing
apparatus should considerable affect the behavior of the measuring
system. These frequent measurements are at the origin of the
so-called Zeno \cite{Misra1977,Peres1980,Itano1990} and anti-Zeno
effects. \cite{Kurizki2000,Kurizki2001} Very recently, Nassar
\cite{Nassar2013a} has proposed a nonlinear logarithmic
Schr\"odinger equation under continuous measurement as a
generalization originally due to Mensky \cite{Mensky1997} and
Bialynicki-Birula and Mycielski \cite{Bialynicki1976}. The
establishment of a  dividing line between the classical and
quantum regimes is one of the main aspects of the measurement
process. \cite{Nassar2013b}

The second application focuses on weak measurements. For this
goal, it is pertinent to analyze the consequences of this GSLE
from a hydrodynamical point of view or following Bohmian mechanics
\cite{salva2013}. This can be carried out by replacing the wave
function $\psi$ (and $\psi^*$) by its real amplitude and real
quantum phase. As it is well known, when conservative systems are
considered, the gradient of this quantum phase gives the momentum
of the particle from the so-called guiding condition (see, for
example, Ref. \cite{salva2013}).  Weak values were proposed by
Aharonov {\it et al.} \cite{Aharonov1988} and are considered to
play a key role in fundamental problems of quantum mechanics (see,
for example, Refs. \cite{Steinberg2011,Bamber2011}). Even more, as
has recently pointed out by Hiley \cite{Hiley2012} in the context
of Bohmian mechanics, these weak values are merely transition
probability amplitudes which, in general, are complex magnitudes.
In particular, for two sequential measurements (one weak and the
other one strong) of complementary observables such as, for
example, momentum and position, the real part provides us the
velocity of the Bohmian particle and the imaginary part the
so-called osmotic velocity due to the presence of a gradient of
the quantum probability. On the contrary, when we are dealing with
open quantum systems, the new guiding condition is issued from our
the previous derived GSLE.

Finally, the third application deals with the generalized
uncertainty principle (GUP), which appears in the context of the
unification of quantum mechanics and general relativity in several
proposals (see \cite{Pedro2013} and references therein). This
principle gives place to deformed commutation relations which are
linear or quadratic in particle momenta. In the linear case, which
corresponds to double special relativity theories
\cite{Magueijo2002}, this fact leads to express the corresponding
dynamics in terms of a gravitational friction within the GLE
framework with a given function $f(x)$. \cite{Pedro2013}
Therefore, the corresponding GSLE can then be easily derived in
terms of this particular function.

Without loss of generality, a one dimensional problem is
considered. For open systems, it is usual to split the total
Hamiltonian into three parts including system, bath and mutual
coupling, in such a way that
\begin{equation}
\label{total}
H=H_{s}+H_{b}+H_{sb},
\end{equation}
where
\begin{equation}
H_{s}=\nobreak \frac{p^{2}}{2m}+V(x)
\end{equation}
stands for the Hamiltonian of the isolated system in presence of a
force field given by the potential $V(x)$;
\begin{equation}
H_{b}=\nobreak \frac{1}{2}\sum_{i} \left(\frac{p_{i}^{2}}{m_{i}}+m_{i}
\omega_{i}^{2}x_{i}^{2} \right)
\end{equation}
is the Hamiltonian for the bath, which acts as a reservoir, and
can be represented as an infinite set of harmonic oscillators; and
\begin{equation}
H_{sb}=\nobreak \sum_{i}\left[\frac{f^{2}(x)d_{i}^{2}}{m_{i}
\omega_{i}^{2}}-2 d_{i}f(x)x_{i} \right]
\end{equation}
expressing the interaction term between the isolated system and
the bath, $d_{i}$ being appropriate coupling constants. The
function $f(x)$ is, in general, a nonlinear function of the system
variable $x$. The term with the square of $f(x)$ gives the
so--called counter term introduced to compensate the
renormalizaton of the potential. Following the standard procedure
where the bath degrees of freedom are eliminated, the equation of
motion for the corresponding system dynamics in the Heisenberg
picture of quantum mechanics is given by the GLE
\cite{Caldeira1981}
%
\begin{eqnarray}
f'\left[x(t)\right]\xi (t) & = & m\ddot x(t) + V'(x) \nonumber \\
& + & m f'\left[x(t)\right]\int_{0}^{t}dt'\alpha(t-t')f'
\left[x(t')\right]\dot x(t') \nonumber \\
\end{eqnarray}
%
where the time--dependent friction (memory kernel) is given by
\begin{equation}
\alpha(t)=\frac{1}{m}\sum_{i}\frac{d_{i}^{2}}{m_{i}\omega^{2}_{i}}
\cos (\omega_{i} t)
\end{equation}
and the external force (noise term) is expressed as
\begin{eqnarray}
\label{noise}
\xi(t)&=&-\sum_{i}d_{i}\left[\left(x_{i}(0)+ \frac{d_{i}}{m_{i}
\omega_{i}^{2}}f(0) \right)\cos (\omega_{i}t)\right] \nonumber \\
&-&d_{i}\sum_{i}\left[\frac{p_{i}(0)}{m_{i}\omega_{i}}\sin(w_{i}t) \right].
\end{eqnarray}
In the Markovian regime, the memory kernel is a $\delta$--function
in time, giving place to an Ohmic dissipation with a
time--independent friction. Within this regime and for a nonlinear
coupling, the corresponding standard Langevin equation reads as
\begin{equation}
\label{eqlang1} m\ddot x(t) + m \alpha \left[f'(x)\right]^{2}\dot
x(t) +V'(x) = f'\left[x(t)\right]\xi (t).
\end{equation}
Notice that the random force is multiplied by the derivative of
the function $f(x)$ giving place to a stochastic process with
multiplicative noise. When the system--bath coupling is linear,
that is, for $f(x)=x$, the standard Langevin equation for additive
noise (that is, when the noise term is not multiplied by any
system function) with Ohmic friction is recovered
\begin{equation}
\label{eqlang2} m\ddot x(t) + m \alpha \dot x(t) +V'(x)= \xi(t) .
\end{equation}

Following the Kostin procedure \cite{Kostin1972}, the generalized
SL equation (that is, for any $f(x)$) can be obtained by writing
first the Schr\"odinger equation as
\begin{equation}
\label{eqSch} i\hbar \frac{\partial \psi}{ \partial
t}=\left[\frac{1}{2m} \left(-i\hbar \frac{\partial}{\partial
x}\right)^2+V(x)+V_{d}+V_{r} \right]\psi,
\end{equation}
where $V_{d}$ and $V_{r}$ are the dissipative and random
potentials to be specified later on. The quantum mechanical
current is defined as
\begin{equation}
J =  \frac{1}{2m}\left[\psi^{*}\left(-i\hbar
\frac{\partial}{\partial x}\right)\psi +\psi\left( -i\hbar
\frac{\partial}{\partial x}\right)\psi^{*}\right]
\end{equation}
\\
Then, from $\partial \psi / \partial t$ and $\partial \psi^{*} /
\partial t$ and Eq. (\ref{eqSch}) we find that
\begin{eqnarray}
\label{JJ}
\frac{d}{dt}\langle x \rangle &=& \int J dx \nonumber \\
\frac{d^{2}}{dt^{2}}\langle x \rangle &=& \int \frac{\partial J}{\partial t} dx.
\end{eqnarray}
where $<.>$ is the expectation value of a given operator.

Now, by performing the same type of averaging into Eq.
(\ref{eqlang1})
\begin{equation}
m\langle \ddot x(t)\rangle + m \alpha \langle
\left[f'(x)\right]^{2} \dot x(t) \rangle +\langle V'(x)\rangle=
\langle f'(x) \xi (t) \rangle,
\end{equation}
and comparing it with Eq. (\ref{eqSch}) and using Eq. (\ref{JJ}),
we can identify terms leading to
\begin{equation}
\int\psi^{*}\left(-\frac{\partial V_{d}}{\partial x}\right)\psi dx
= \alpha m \int f'(x)^2 J dx.
\end{equation}
Thus, the damping potential is a functional of the wave function
and can be expressed as
\begin{equation}
\label{Vd} V_{d}\left[\psi,\psi^{*},f \right]=-m\alpha \int
\frac{\tilde J} {\psi\psi^{*}} dx,
\end{equation}
where the new quantum mechanical current is now defined as
\begin{equation}
\tilde J  \equiv f'(x)^{2} J
\end{equation}
which is coupling--dependent. On the other hand, the generalized
random potential due to the heat bath corresponding to the random
force $\xi(t)$ can be written as
\begin{equation}
\label{random}
V_{r}=-f(x)\xi(t).
\end{equation}
Finally, the corresponding  GSLE can be then expressed by
\begin{widetext}
\begin{equation}
\label{GSLE} i\hbar \frac{\partial \psi}{ \partial
t}=\left[\frac{1}{2m}\left(-i\hbar \frac{\partial}{\partial
x}\right)^2+V(x)- m\alpha \int \frac{\tilde J}{\psi\psi^{*}}
dx-f(x)\xi(t) - W(t) \right]\psi.
\end{equation}
\end{widetext}
where $W(t) = \langle V_d \rangle$ arises from the requirement
that the integration of Eq. (\ref{Vd}) with respect to $x$ must be
equal to the expectation values of the kinetic and potential
energies through the total Hamiltonian. As mentioned by Kostin
\cite{Kostin1972}, this term can be removed from Eq. (\ref{GSLE})
by introducing the transformation of the wave function
$\psi(x,t)=e^{i\theta(t)}\phi(x,t)$. We also note that when the
coupling function $f$ is assumed to be linear in the system
variable, Eq. (\ref{GSLE}) reduces to the standard SL or Kostin
equation \cite{Kostin1972}.

A new and straightforward generalization of Eq. (\ref{GSLE}) is
when continuous measurement is considered. As has been recently
shown, this process can also be described by a nonlinear
logarithmic Schr\"odinger equation \cite{Nassar2013a,Nassar2013b}.
Thus, we have
\begin{widetext}
\begin{equation}
\label{CM} i\hbar \frac{\partial \psi}{ \partial
t}=\left[\frac{1}{2m}\left(-i\hbar \frac{\partial}{\partial
x}\right)^2+V(x)- m\alpha \int \frac{\tilde J}{\psi\psi^{*}}
dx-f(x)\xi(t) - W(t) + W_{\kappa} (x,t)\right]\psi.
\end{equation}
\end{widetext}
where $W_{\kappa}(x,t)= - i \hbar \kappa [ \ln |\psi(x,t)|^2 -
\langle \ln |\psi(x,t)|^2 \rangle ]$ and $\kappa$ gives the
resolution of the continuous measurement. These two basic
decoherence mechanisms are thus put on equal footing. This more
general equation should be applicable to the Zeno and anti-Zeno
effects in presence of an environment displaying nonlinear
dissipation and, in general, to the so-called environment induced
decoherence. \cite{Piilo2006}

Our next goal is to propose an equation governing the weak
measurement process in presence of a heat bath. For this purpose,
instead of dealing with the two fields $\psi$ and $\psi^*$, Eq.
(\ref{GSLE}) can be written in terms of the hydrodynamical or
quantum trajectory formulation by expressing the wave function in
polar form with a real amplitude $A(x,t)$ and real phase $S(x,t)$
as
\begin{equation}
\label{fdo} \psi(x,t)=A(x,t)e^{i S(x,t)/\hbar}.
\end{equation}
Then, Eq. (\ref{eqSch}) can now be split into a system of two
coupled equations,
\begin{eqnarray}
\label{eqphase}
\frac{\partial S}{\partial t}&=&-\frac{1}{2m}\left(\frac{\partial S}
{\partial x}\right)^{2}-(V+V_d+V_r+Q) \\
\label{eqamplitude} \frac{\partial A^{2}}{\partial
t}&=&-\frac{1}{m}\frac{\partial S} {\partial x}\frac{\partial
A^{2}}{\partial x}- \frac{A^{2}}{m}\frac{\partial^{2}S}{\partial
x^{2}},
\end{eqnarray}
where
\begin{equation}
Q(x,t)=-\frac{\hbar^{2}}{2mA}\frac{\partial^{2}A}{\partial x^{2}}
\end{equation}
is the quantum potential. Note that the first and second equations
are the quantum Hamilton-Jacobi and continuity equations,
respectively. Moreover, the current density is expressed in this
formalism as
\begin{equation}
\label{JB} J = \frac{\psi\psi^{*}}{m}\frac{\partial S} {\partial
x} .
\end{equation}

As is known, the gradient of the wave function phase is associated
with the trajectory momentum by means of $p(x,t)=\partial
S(x,t)/\partial x$ (the guiding condition). By differentiating Eq.
(\ref{eqphase}), the time evolution of $p(x,t)$  is given by
\begin{equation}
\frac{\partial p}{\partial t}=-\frac{p}{m}\frac{\partial
p}{\partial x} -\frac{\partial}{\partial
x}\left(V+V_d+V_r+Q\right).
\end{equation}

According to Eq. (\ref{eqlang1}), which can be interpreted in
terms of the Lagrangian framework of hydrodynamics, the
corresponding quantum Newton-Langevin equation including the
dissipative and random sources can be expressed as
\begin{eqnarray}
\label{eqintegral} \frac{\partial p}{\partial t} & = &
-\frac{p}{m}\frac{\partial p}{\partial x}
-\frac{\partial}{\partial x}\left(V+Q\right) \nonumber \\
& - & \alpha f'(x)^{2}p-f'(x) \xi(t).
\end{eqnarray}
Then, by integrating Eq. (\ref{eqintegral}) with respect to $x$,
we obtain
\begin{eqnarray}
\label{St}
 -\frac{\partial S}{\partial t} & = &
\frac{p^{2}}{2m}+V+Q+ \alpha\int f'(x)^{2} p dx
\nonumber \\
& + & f(x)\xi(t)+C(t),
\end{eqnarray}
where $C(t)$ is an arbitrary time function resulting from the
space integration to be specified later on. This equation gives
the evolution of the wave function phase in presence of damping
which corresponds to a generalized Caldeira--Leggett coupling.
Clearly, the random potential is also expressed in this case by
Eq. (\ref{random}).

Moreover, the partial integration in Eq. (\ref{St}) leads to
\begin{equation}
\int \left( \frac{d f}{d x}\right)^{2} p dx = \left( \frac{d f}{d
x}\right)^{2} S -2\int S \frac{d f}{d x} \frac{d^{2} f}{d x^{2}}
dx,
\end{equation}
which gives place naturally to the coupling--dependent phase
\begin{equation}
\label{newphase} \tilde S \equiv \left( \frac{d f}{d x}\right)^{2}
S -2\int S \frac{d f}{d x} \frac{d^{2} f}{d x^{2}} dx.
\end{equation}
which can be straightforwardly expressed as
\begin{equation}
\tilde S =  m \int \frac{\tilde J}{\psi\psi^{*}} dx
\end{equation}
with
\begin{equation}
\label{eqnewphase} V_{d}\left[\psi,\psi^{*},f\right] = - \alpha
\tilde S   .
\end{equation}
Importantly, Eq. (\ref{eqnewphase}) includes, as a special case,
the Bohmian version of the Kostin equation when the coupling is
linear. In this case, $\tilde J\rightarrow J$ and
$V_{d}\left[\psi,\psi^{*},f\right]\rightarrow
V_{d}\left[\psi,\psi^{*},x\right]=-\alpha S$, which is the
dissipative potential expressed within the Bohmian formalism
\cite{Razavi2005,Nassar2013b,Gara2013}.

The constant of integration $C(t)$ can be defined in such a way
that the overall phase of the wave function should not affect its
evolution. This requirement is satisfied when $C(t)=-\alpha
\langle \tilde S\rangle$. Therefore, $C(t) \equiv W(t)$ and for a
general non--linear coupling, the GSLE or Kostin equation can be
rewritten in terms of the modified quantum action, $\tilde S$, as
\begin{widetext}
\begin{equation}
\label{GSLE-B} i\hbar \frac{\partial \psi}{ \partial t}  =
\left[\frac{1}{2m} \left(-i\hbar \frac{\partial}{\partial
x}\right)^2+V(x) - \alpha \left(\tilde S -\langle \tilde
S\rangle\right) - f(x) \xi (t) \right]\psi    .
\end{equation}
\end{widetext}
Thus, the corresponding generalized Hamilton-Jacobi equation for
the nonlinear dissipation is given by Eq. (\ref{eqphase}) with
$V_d$ given by Eq. (\ref{eqnewphase}) and $V_r$ by Eq.
(\ref{random}). Notice, however, that the continuity equation is
the same expressed by Eq. (\ref{eqamplitude}). As before, the term
$W_{\kappa}$ could also be added for describing the continuous
measurement process.

After Hiley \cite{Hiley2012}, the sequential measurement of the
momentum and position of a particle of mass $m$ is given by the
weak value or transition probability amplitude
\begin{equation}
\frac{\langle x | p | \psi \rangle}{\langle x | \psi \rangle} =
\frac{\partial S}{\partial x} - \frac{i}{2 A^2} \frac{\partial
A^2}{\partial x}
\end{equation}
where $\psi$ obeys the standard (conservative and linear)
time-dependent Schr\"odinger equation and is expressed in polar
form according to Eq. (\ref{fdo}). The real part is the Bohmian
velocity of the particle and the imaginary part the so-called
osmotic velocity due to the presence of the gradient of the
quantum probability. In this context, the same weak value is
obtained in presence of an environment but now $\psi$ is governed
by Eq. (\ref{GSLE}) for linear and/or nonlinear dissipation.

Finally, let us apply this formalism to the GUP. As shown recently
\cite{Pedro2013}, the deformed commutation relations $[x,p]=i\hbar
\left(1-\frac{\gamma}{m_{p}c} p \right)$, where $\gamma$ is the
dimensionless GUP parameter and $m_{p}$ is the Planck mass, can be
expressed as $[x,p]=i\hbar \left(1-\frac{\alpha}{2} p \right)$
where $\alpha\equiv \gamma/m_{p}c$ is a friction coefficient.
Moreover, this commutation relation leads to a GLE with a
position--dependent coupling that turns out to be
$f(x)=\int_{0}^{x}\sqrt{V'(y)}dy$. In this case, working within
the Bohmian formalism we arrive at the following damping potential
for the GUP,
\begin{equation}
V_{d}=-\alpha \tilde S= -\frac{2\gamma}{m_{p}c}p\, V(x).
\end{equation}
Therefore, the corresponding GSLE can be written as
\begin{widetext}
\begin{equation}
\label{GUP-eq} i\hbar \frac{\partial \psi}{ \partial t}  =
\left[\frac{1}{2m} \left(-i\hbar \frac{\partial}{\partial
x}\right)^2+V(x)\left(1- \frac{2\gamma}{m_{p}c} p \right) - \frac{2\gamma}{m_{p}c}\langle p\, V(x)\rangle\right]\psi    .
\end{equation}
\end{widetext}
Notice, that, although Eq. (\ref{GUP-eq}) predicts the correct Langevin--like equation derived from the
deformed commutation relations (Eq. (6) of ref. \cite{Pedro2013}), the correct
dynamics for the phase space variables can not be obtained from it. This situation
is similar to that found within the well known Caldirola--Kanai effective Hamiltonian approach
 \cite{salva2013}.

In summary, along this work we have derived a general SL equation
valid for quantum processes in presence of nonlinear friction and
a heat bath. This equation is reduced to the Kostin equation for
linear friction. Afterwards, we have focused on the measurement
process and studied two main topics in this context, the
continuous measurement process and weak measurements. In the first
issue, we have extended the GSLE to include frequent observations
and proposed the equation governing the corresponding dynamics.
This equation should be applicable to the Zeno and anti-Zeno
effects for open systems. In the second issue, we have rewritten
the GSLE within the Bohmian formalism and expressed the guiding
condition for obtaining the corresponding stochastic quantum
trajectories. Afterwards, weak values have been discussed in terms
of these quantum trajectories when a sequential measurement of
momentum and position of a mass particle are carried out. Finally,
in the context of the generalized uncertanty principle \cite{Pedro2013}, the corresponding GLSE
has also been derived.

This work has been funded by the MICINN (Spain) through Grant No.
FIS2011-29596-C02-C01. S. M.-A. acknowledges the COST
Action MP1006 entitled "Fundamental Problems in Quantum Physics"

\end{document}